\documentclass[11pt]{article}

\usepackage{amsmath,amssymb,amsthm}
\usepackage[numbers]{natbib}
\usepackage{graphicx}
\usepackage{booktabs}
\usepackage{geometry}
\usepackage{bm}
\usepackage{enumitem}
\usepackage{caption}
\usepackage{subcaption}
\usepackage{float}
\usepackage{xcolor}
\usepackage{tikz}
\usepackage{array}
\usepackage{xcolor}

\usepackage[numbers]{natbib}
\definecolor{deepblue}{RGB}{0,70,140}
\definecolor{tealgreen}{RGB}{0,120,110}
\definecolor{darkred}{RGB}{150,40,40}
\usepackage[
    colorlinks=true,
    linkcolor=darkred,
    citecolor=tealgreen,
    urlcolor=deepblue
]{hyperref}

\usetikzlibrary{arrows.meta,positioning,fit,shapes.geometric,calc}

\usepackage{titling}
\usepackage{array}
\preauthor{\begin{center}
    \large \lineskip .75em%
        \begin{tabular}[t]{>{\centering\arraybackslash}p{.45\textwidth}}}
        \postauthor{\end{tabular}\par\end{center}}
\makeatletter
\renewcommand\and{
  \end{tabular}%
  \hfill
  \begin{tabular}[t]{>{\centering\arraybackslash}p{.45\textwidth}}}
\makeatother

\renewcommand{\toprule}{\hline\hline}
\renewcommand{\midrule}{\hline}
\renewcommand{\bottomrule}{\hline}

\usepackage{mathtools}
\newcommand{\defeq}{\vcentcolon=}

\title{%
Loop Current Extension as an Effective Delayed Dynamical System
}

\author{%
Francisco J. Beron-Vera\thanks{Corresponding author.}\\
Department of Atmospheric Sciences\\
Rosenstiel School of Marine, Atmospheric, and Earth Science\\
University of Miami\\
Miami, Florida, USA\\
\href{mailto:fberon@miami.edu}{\texttt{fberon@miami.edu}}
\and
Maria J. Olascoaga\\
Department of Ocean Sciences\\
Rosenstiel School of Marine, Atmospheric, and Earth Science\\
University of Miami\\
Miami, Florida, USA\\
\href{mailto:jolascoaga@miami.edu}{\texttt{jolascoaga@miami.edu}}
\and
Philippe Miron\\
Department of Ocean Sciences\\
Rosenstiel School of Marine, Atmospheric, and Earth Science\\
University of Miami\\
Miami, Florida, USA\\
\href{mailto:pmiron@miami.edu}{\texttt{pmiron@miami.edu}}
}

\date{Started: May 28, 2026.  This version: \today.}

\begin{document}

\maketitle

\begin{abstract}
The Loop Current is the dominant circulation feature of the Gulf of Mexico and exhibits pronounced variability associated with northward extension, retraction, and eddy shedding. Despite decades of study, the extent to which this variability admits a reduced dynamical description remains unclear.

We investigate this question using delayed-coordinate representations constructed from satellite-altimetry observations of Loop Current extension. Ridge regression, multilayer perceptron forecasting, and Sparse Identification of Nonlinear Dynamics (SINDy) are applied to learn delayed evolution maps from the extension time series. Forecast skill consistently exceeds persistence at lead times of 30--90 days while requiring only a small number of delayed coordinates. Ridge regression reveals saturation with delayed-state dimension, indicating that much of the predictive information is contained within a compact representation. Neural-network forecasts provide modest additional improvements, while delayed SINDy identifies sparse evolution maps involving intraseasonal memory scales, from approximately two weeks to a few months, that remain stable under recursive iteration. Physical diagnostics associated with Yucatan Channel inflow, Florida Straits outflow, gateway geometry, and northern Caribbean vorticity contain predictive information but do not provide additional independent state information once the delayed Loop Current state is included.

These results support the interpretation of Loop Current extension as an observable evolving on an effective low-dimensional delayed dynamical system. A substantial fraction of the predictable variability can be reconstructed from a small number of delayed observations and represented through compact delayed evolution maps.
\end{abstract}

\noindent\textbf{Satellite-altimetry observations reveal that Loop Current extension variability can be represented by a compact delayed state. Forecast skill saturates with only a few delayed coordinates, sparse delayed maps remain stable under recursive iteration, and physical diagnostics provide limited additional information beyond the reconstructed state. These results support an effective low-dimensional delayed description of Loop Current evolution.}

\tableofcontents

\section{Introduction}

The Loop Current is the dominant circulation feature of the Gulf of Mexico \cite{Vukovich-95, Oey-etal-13}. Its variability is characterized by cycles of northward extension, retraction, and the shedding of large anticyclonic rings that influence transport, mixing, and predictability throughout the basin. Despite decades of observational and modeling studies, it remains unclear to what extent this variability can be represented by a reduced dynamical description.

Previous studies have shown that important aspects of Loop Current variability can be related to a small number of geometric descriptors. In particular, relationships between Loop Current retreat latitude and subsequent eddy-shedding intervals suggest that a reduced description may capture a substantial fraction of the observed variability \cite{Sturges-Leben-00, Lugo-Leben-10}. Machine-learning approaches have also been used to identify recurring Loop Current states from satellite-altimetry observations through self-organizing maps, providing low-dimensional representations of Loop Current variability \cite{Weisberg-etal-17}. These results motivate the broader question of whether Loop Current variability admits a compact dynamical representation. In this work, this variability is described using the Loop Current extension \(L(t)\), a scalar observable measuring the northward extent of the Loop Current front.

Takens' embedding theorem states that if a dynamical system evolves on a finite-dimensional attractor, then the geometry of this attractor can generically be reconstructed from sufficiently many delayed observations of a single scalar observable \cite{Takens-81}. This suggests that the history of \(L(t)\) may contain information about the effective state controlling Loop Current evolution when represented through appropriate delayed coordinates.

The full ocean circulation state, including velocity, pressure, density, and other fields, is extremely high dimensional. The delayed representation of \(L(t)\) is not intended to recover this complete state, but to provide coordinates for a reduced description of the observed Loop Current variability. This motivates the search for compact delayed states that contain sufficient information to define predictive evolution maps.

The objective of this work is to determine whether such a representation exists for Loop Current extension variability and to investigate how this reconstructed state relates to physically motivated diagnostics. Four questions are considered:
\begin{enumerate}[label=(\roman*),nosep]
    \item whether a compact delayed state captures a substantial fraction of the predictable variability;
    \item whether nonlinear forecasting provides significant advantages beyond such a representation;
    \item whether the resulting dynamics can be represented through sparse delayed evolution maps; and
    \item whether transport, geometric, and mesoscale diagnostics provide additional coordinates for the effective state.
\end{enumerate}

To address these questions, a scalar observable describing the northward extent of the Loop Current is extracted from satellite-altimetry observations and analyzed using delayed-coordinate reconstruction. Ridge regression \cite{Hoerl-Kennard-70} is employed to assess how much predictive information is contained within delayed-state representations with different numbers of coordinates and memory lengths. A multilayer perceptron \cite{Rumelhart-etal-86} is then used to determine whether flexible nonlinear forecasting models provide substantial gains beyond the delayed-state representation itself. The choice of a multilayer perceptron keeps the delayed state explicit, allowing the memory structure to be examined directly and compared with the ridge and sparse-model results. Sparse Identification of Nonlinear Dynamics (SINDy) \cite{Brunton-etal-16} is used to investigate whether the resulting dynamics admit a parsimonious symbolic representation. Finally, transport, geometric, and mesoscale diagnostics associated with the Yucatan Channel, Florida Straits, and northern Caribbean Sea are examined to determine whether they provide additional state information beyond the delayed Loop Current history.

The emphasis throughout is on identifying the delayed-state structure underlying Loop Current variability rather than maximizing forecast skill through increasingly complex forecasting architectures. The delayed coordinates are therefore prescribed explicitly and used consistently across the ridge, multilayer-perceptron, and SINDy analyses.

\section{Data and Loop Current extension}

Operational studies have often used the 17-cm sea-level-anomaly contour introduced by Leben \cite{Leben-05}, chosen because it closely tracks the edge of the high-velocity Loop Current core. Because the present analysis is based on absolute dynamic topography (ADT), the corresponding frontal contour is calibrated directly from the ADT field rather than prescribed from an SLA threshold. Since surface geostrophic flow is proportional to \(\nabla^\perp\eta\), the strongest geostrophic transition is expected to coincide with regions of large \(|\nabla\eta|\).

The ADT level defining the front is determined through a geometric calibration. A range of candidate values \(\eta_0\) is considered. For each value, the corresponding Loop Current ADT contour
\begin{equation}
    \Gamma_{\eta_0}(t_n)
    =
    \{
    (\lambda,\vartheta):
    \eta(\lambda,\vartheta,t_n)=\eta_0
    \}
\end{equation}
is extracted at each time \(t_n\). The selected threshold is the contour level that maximizes the average ADT gradient magnitude along these candidate fronts over the observational record:
\begin{equation}
    \eta_c
    =
    \arg\max_{\eta_0}
    \frac{1}{N}
    \sum_{n=1}^{N}
    \frac{\int_{\Gamma_{\eta_0}(t_n)}
    |\nabla\eta(\lambda,\vartheta,t_n)|\,ds}
    {\int_{\Gamma_{\eta_0}(t_n)}ds}.
\end{equation}
Here \(ds\) denotes arclength along the contour. The ratio of line integrals gives the mean ADT gradient magnitude along the candidate front at a given time, while the summation averages this quantity over the \(N\) satellite snapshots used for calibration. This procedure gives
\begin{equation}
    \eta_c\defeq0.55~\mathrm{m}.
\end{equation}
The value of \(\eta_c\) is then held fixed for all subsequent analysis and is not adjusted to optimize the delayed-coordinate forecasts. This choice provides a consistent observable whose variability reflects changes in Loop Current geometry rather than changes in the contour-selection criterion. Fully Lagrangian boundary definitions, such as shearless transport barriers associated with the Loop Current jet \cite{Farazmand-etal-14, Beron-etal-20-Chaos}, provide objective descriptions of coherent transport structures. Here, the calibrated ADT contour is used as a reproducible Eulerian measure of Loop Current extension, keeping the focus on the delayed evolution of the observed large-scale variability rather than on boundary extraction.

\begin{figure}[t!]
    \centering
    \includegraphics[width=\textwidth]{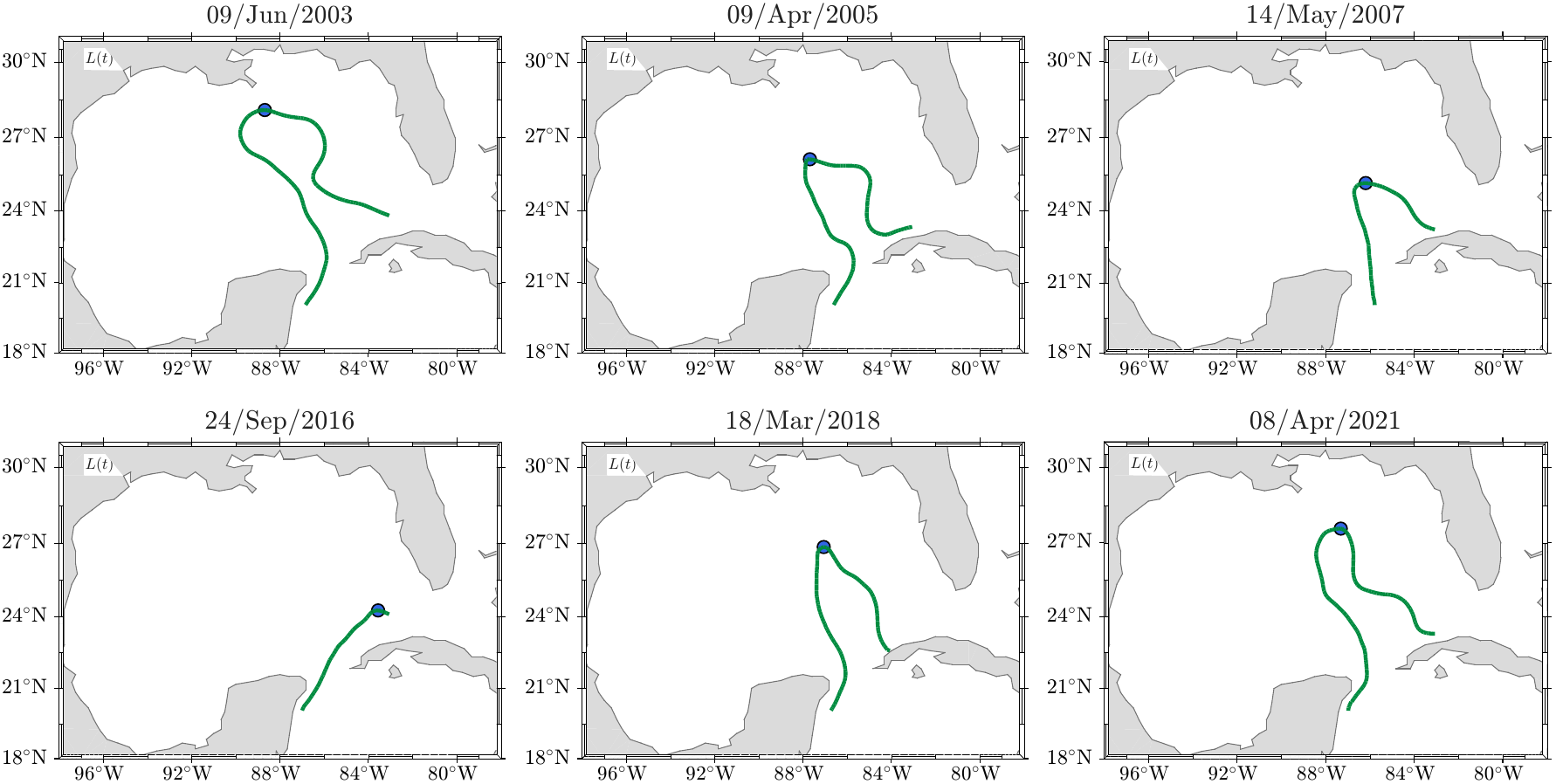}
    \caption{
        Representative Loop Current frontal-contour snapshots. The green curve is the selected Gulf-side segment of the ADT contour \(\eta=\eta_c\), with \(\eta_c=0.55~\mathrm{m}\). The blue marker indicates the northernmost point of the selected contour segment and defines the Loop Current extension \(L(t)\). The sequence illustrates retracted, intermediate, and extended Loop Current configurations.
    }
    \label{fig:lc-frontal-contour-sequence}
\end{figure}

The Loop Current enters the Gulf of Mexico through the Yucatan Channel and exits through the Florida Straits. Variability in the position and geometry of the Loop Current front reflects the cumulative effects of these inflow and outflow pathways together with the internal evolution of the current and associated eddy-shedding events. The selected ADT contour therefore provides a compact geometric description of the large-scale Loop Current state. In representative examples shown in Fig.~\ref{fig:lc-frontal-contour-sequence}, the selected contour segment captures the evolving Loop Current front across retracted, intermediate, and extended states.

Let
\begin{equation}
    \Gamma_{\eta_c}(t)
    \defeq
    \left\{
        (\lambda,\vartheta)
        :
        \eta(\lambda,\vartheta,t)=\eta_c,
        \;
        \lambda\le -83^\circ,
        \;
        \vartheta\ge 20^\circ
    \right\}.
\end{equation}
The restrictions
\(
\lambda\le -83^\circ
\)
and
\(
\vartheta\ge 20^\circ
\)
exclude the Florida Current and adjacent Atlantic continuation of the same level set, thereby isolating the Loop Current front within the Gulf of Mexico and preventing the extension metric from being influenced by high-latitude points along the outflow through the Florida Straits. The set \(\Gamma_{\eta_c}(t)\) may contain multiple connected contour segments. The segment corresponding to the Loop Current front is denoted by \(\Gamma_{\eta_c}^\text{LC}(t)\). The Loop Current extension is then defined by
\begin{equation}
    L(t)
    \defeq
    \max
    \left\{
        \vartheta
        :
        (\lambda,\vartheta)\in\Gamma_{\eta_c}^\text{LC}(t)
    \right\},
\end{equation}
that is, the northernmost latitude attained by the selected Loop Current frontal contour.

Verification of the contour database confirmed that the stored extension time series is exactly this quantity. The resulting record converts the evolving Loop Current geometry into a single scalar observable suitable for delayed-state reconstruction and dynamical-system analysis.

\section{Delayed-state representation of Loop Current extension}

Forecast performance is evaluated relative to a persistence forecast. For a forecast horizon \(\Delta\), persistence predicts
\begin{equation}
    \widehat L_{\rm pers}(t+\Delta)
    =
    L(t).
\end{equation}
Let \({\rm RMSE}_{\rm model}\) denote the root-mean-square forecast error of a given model and \({\rm RMSE}_{\rm pers}\) the corresponding persistence error. Forecast skill is quantified using the normalized RMSE ratio
\begin{equation}
    R
    \defeq
    \frac{{\rm RMSE}_{\rm model}}
         {{\rm RMSE}_{\rm pers}}.
\end{equation}
Values \(R<1\) indicate improvement over persistence, while \(R=1\) corresponds to persistence.

Unless otherwise stated, all reported forecast ratios are computed on a held-out test interval not used for model fitting or parameter selection. Direct forecasts are evaluated using observed delayed coordinates as inputs to the learned forecasting maps. Model selection is performed using only the training and validation data according to the procedure appropriate for each method.

\subsection{Delayed-state structure}

The delayed-state representation was first examined using ridge regression, a regularized linear forecasting model. The objective of this step is not to identify a sparse evolution law, but rather to determine whether candidate delayed coordinates contain sufficient information to predict the future evolution of the LC extension. Given a delayed-coordinate vector
\begin{equation}
    \mathbf z_n(\mathcal T)
    =
    \bigl(
    L(t_n-\tau_1),
    L(t_n-\tau_2),
    \ldots,
    L(t_n-\tau_m)
    \bigr),
\end{equation}
where
\begin{equation}
    \mathcal T
    =
    \{\tau_1,\tau_2,\ldots,\tau_m\}
\end{equation}
denotes the selected lag set, ridge regression approximates the delayed evolution map by the linear projection
\begin{equation}
    L(t_n+\Delta)
    \approx
    \alpha_0
    +
    \boldsymbol{\alpha}^{\top}\mathbf z_n(\mathcal T).
\end{equation}
The intercept and regression coefficients are obtained from the regularized least-squares problem
\begin{equation}
    (\alpha_0,\boldsymbol{\alpha})
    =
    \arg\min_{\beta_0,\boldsymbol{\beta}}
    \sum_n
    \Big(
    L(t_n+\Delta)
    -
    \big(
    \beta_0
    +
    \boldsymbol{\beta}^{\top}\mathbf z_n(\mathcal T)
    \big)
    \Big)^2
    +
    \lambda\|\boldsymbol{\beta}\|_2^2,
\end{equation}
where $\lambda>0$ is a regularization parameter \cite{Hoerl-Kennard-70}. The $L^2$ penalty reduces sensitivity to nearly redundant predictors by shrinking coefficient magnitudes without imposing sparsity. This property is particularly useful for delayed-coordinate representations, where neighboring lags of a slowly evolving observable are expected to be strongly correlated. Ridge regression therefore provides a stable measure of the predictive information contained in a candidate delayed state before introducing nonlinear or sparse models.

\begin{figure}[t!]
    \centering
    \includegraphics[width=.75\textwidth]{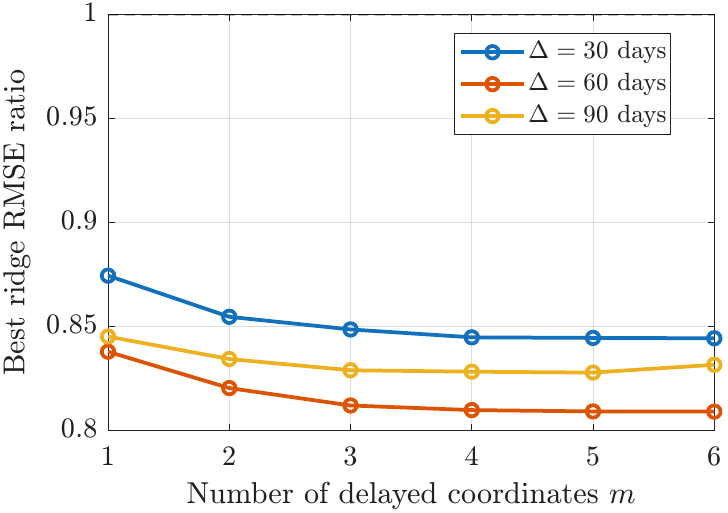}
    \caption{
        Ridge forecast skill as a function of delayed-state dimension \(m\) for 30-, 60-, and 90-day forecasts. For each \(m\), the plotted value corresponds to the best-performing combination of \(m\) delayed coordinates selected from the candidate lag set.
    }
    \label{fig:ridge_dimension}
\end{figure}

The ridge analysis showed that forecast skill improves rapidly as delayed coordinates are added and then saturates after only a few delays (Fig.~\ref{fig:ridge_dimension}). Here, the delayed-state dimension \(m\) denotes the number of delayed coordinates retained in the state, not a uniformly spaced delay embedding. For each value of \(m\), all combinations of \(m\) delays were tested from the candidate lag library and the best-performing set was selected. The selected delays should therefore be interpreted as preferred memory scales rather than as a unique delay spacing.

The best-performing lag combinations consistently included the present Loop Current extension together with a small number of delayed observations. Repeating the analysis using a finer lag library showed that the precise selected delays shift within neighboring windows, while the main result remains unchanged: forecast skill saturates after approximately three to five coordinates, with dominant memory scales of approximately 15--45 days and additional contributions near 65--75 days for some forecast horizons.

\subsection{Predictive reconstruction}

The ridge analysis indicates that a small number of delayed observations captures most of the predictive information available in the Loop Current extension time series. The next question is whether nonlinear interactions among the delayed coordinates contain additional predictive information. To investigate this possibility, multilayer perceptron (MLP) forecasting models \cite{Rumelhart-etal-86} were trained using delayed-coordinate representations of the extension.

Given a delayed-state vector $\mathbf z_n(\mathcal T)$, the MLP approximates the forecast map
\begin{equation}
    \widehat L(t_n+\Delta)
    =
    F_{\theta}(\mathbf z_n(\mathcal T)),
\end{equation}
where $F_{\theta}$ is a nonlinear function represented by a feedforward neural network with parameters $\theta$. Each hidden layer applies an affine transformation followed by a nonlinear activation,
\begin{equation}
    \mathbf h^{(\ell+1)}
    =
    \sigma\big(
        W^{(\ell)}
        \mathbf h^{(\ell)}
        +
        \mathbf b^{(\ell)}
    \big),
\end{equation}
with input $\mathbf h^{(0)}=\mathbf z_n(\mathcal T)$, weight matrices $W^{(\ell)}$, bias vectors $\mathbf b^{(\ell)}$, and activation function $\sigma$. The network parameters are obtained by minimizing the mean-squared forecast error
\begin{equation}
    \min_{\theta}
    \sum_n
    \big(
        L(t_n+\Delta)
        -
        F_{\theta}(\mathbf z_n(\mathcal T))
    \big)^2.
\end{equation}
The available time series was divided chronologically into training, validation, and test intervals containing 70\%, 15\%, and 15\% of the data, respectively. The validation interval was used for early stopping and model selection among different random initializations, while all reported forecast ratios were computed on the independent test interval.

Unlike ridge regression, which assumes a linear dependence on the delayed coordinates, the MLP can represent nonlinear interactions among the components of the delayed state. The purpose of the MLP analysis is therefore to test whether additional nonlinear structure in the delayed evolution map provides predictive information beyond that already contained in the reconstructed state.

Forecast ratios were substantially below persistence when delayed observations of the Loop Current extension were used as predictors. Based on the ridge analysis, the nonlinear forecasts were evaluated using compact lag sets representative of the identified memory windows rather than attempting to optimize individual delay values. This allows the MLP experiments to test whether nonlinear interactions among the reconstructed coordinates provide additional predictive information beyond the delayed-state structure itself.

Forecast performance for the best anomaly-delayed MLP forecasts is summarized in Table~\ref{tab:mlp_forecast_skill}. The most successful forecasts were obtained using compact delayed states containing only a few coordinates. For the 30- and 60-day horizons the preferred states involve memory on approximately two-week and one-to-two-month time scales, consistent with the ridge results. At the 90-day horizon the selected state includes a longer memory coordinate, although the differences among compact delayed representations remain small.

\begin{table}[t!]
    \centering
    \begin{tabular}{lcc}
        \toprule
        Forecast horizon & RMSE ratio \(R\) & Preferred lag set \\
        \midrule
        30 days & 0.832 & $[0,14,45]$ days \\
        60 days & 0.791 & $[0,14,45]$ days \\
        90 days & 0.804 & $[0,30,90]$ days \\
        \bottomrule
    \end{tabular}
    \caption{
      Forecast skill of MLP models using representative compact delayed-state representations identified from the ridge memory-scale analysis. Values show forecast error relative to persistence.
    }
    \label{tab:mlp_forecast_skill}
\end{table}

\begin{figure}[t!]
    \centering
    \includegraphics[width=.75\textwidth]{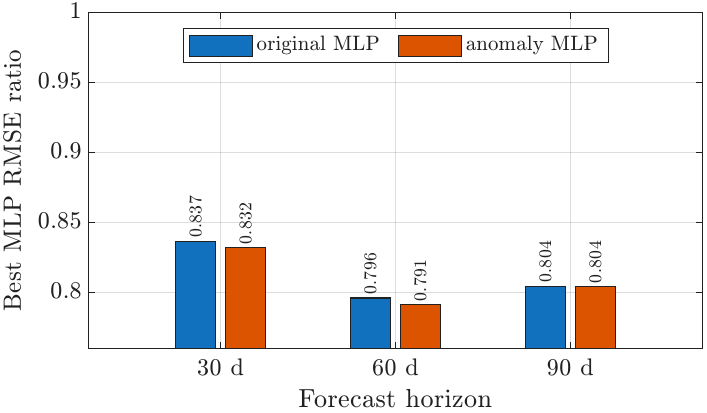}
    \caption{
        Comparison of forecast skill obtained using the original Loop Current extension time series and extension anomalies obtained by removing the climatological annual cycle.
    }
    \label{fig:lc_raw_vs_anomaly}
\end{figure}

To assess the extent to which forecast skill derives from delayed-state information rather than from the mean seasonal evolution, an additional set of experiments was performed using extension anomalies. The anomaly series was obtained by subtracting the climatological annual cycle from the original Loop Current extension time series, thereby removing the mean seasonal variability and emphasizing departures from the typical seasonal state. Figure~\ref{fig:lc_raw_vs_anomaly} compares forecasts obtained from the original extension time series and from the anomaly series. Training on anomalies produces only marginal changes in forecast skill. This indicates that predictive skill is not primarily driven by the climatological seasonal cycle. Instead, most of the predictive information is contained in the delayed evolution of the Loop Current extension itself, and removing the mean seasonal variability provides little additional benefit.

The delayed-MLP forecasts outperform persistence using compact delayed states with only a few coordinates. Their improvement relative to ridge regression is measurable but modest, indicating that much of the predictable variability is already captured by the delayed coordinates. The nonlinear model mainly refines the forecast within this reconstructed state space, while the delayed-state representation provides the dominant source of predictive skill.

\subsection{Sparse delayed dynamical maps}

Having identified compact delayed embeddings with predictive skill, the next question is whether the corresponding dynamics admit a parsimonious delayed representation. To address this question, Sparse Identification of Nonlinear Dynamics (SINDy) \cite{Brunton-etal-16} was applied directly to delayed-coordinate representations of the Loop Current extension. SINDy seeks a sparse approximation of the forecast map by expressing the future state as a linear combination of candidate basis functions and then eliminating unnecessary terms through sparse regression.

Given a delayed-state vector $\mathbf z_n(\mathcal T)$, delayed SINDy seeks a sparse approximation of the delayed evolution map
\begin{equation}
    \widehat L(t_n+\Delta)
    =
    F\!\left(\mathbf z_n(\mathcal T)\right)
    =
    \sum_{k=1}^{p}
    \xi_k\,
    \Theta_k\!\left(\mathbf z_n(\mathcal T)\right),
\end{equation}
where $\Theta_k$ are candidate basis functions and $\xi_k$ are coefficients selected from a sparse library. The candidate library consisted of polynomial functions of the delayed coordinates up to degree three, including constant, linear, quadratic, and cubic terms. For example, if the selected lag set is $\mathcal T=\{\tau_1,\tau_2,\tau_3\}$, then
\begin{equation}
    \mathbf z_n(\mathcal T)
    =
    \bigl(
    L(t_n-\tau_1),
    L(t_n-\tau_2),
    L(t_n-\tau_3)
    \bigr).
\end{equation}
The corresponding polynomial library contains functions such as
\begin{equation}
\begin{aligned}
    &1,\quad
    L(t_n-\tau_1),\quad
    L(t_n-\tau_2),\quad
    L(t_n-\tau_3),\\
    &L(t_n-\tau_1)L(t_n-\tau_2),\quad
    L(t_n-\tau_1)^2,\\
    &L(t_n-\tau_1)L(t_n-\tau_2)L(t_n-\tau_3),
\end{aligned}
\end{equation}
together with all other polynomial terms of degree at most three in the delayed coordinates. Sparse coefficients are identified using Sequentially Thresholded Least Squares (STLSQ) \cite{Brunton-etal-16}, which iteratively removes coefficients whose magnitude falls below a prescribed threshold and recomputes the remaining coefficients by least squares. Since the delayed coordinates have already been selected through the ridge analysis, the sparsification step is not used to determine the memory structure, but rather to identify a parsimonious nonlinear map on the reconstructed delayed state. The resulting approximation retains only a small subset of active terms, yielding an interpretable symbolic representation of the delayed evolution map.

The delayed-SINDy forecasts are summarized in Table~\ref{tab:sindy_results}. Following the ridge analysis, SINDy was applied to compact delayed states representative of the identified memory windows in order to determine whether these reconstructed states admit sparse evolution laws. The selected models are consistent with the ridge and delayed-MLP results, involving the present LC extension together with memory on approximately two-week to one-to-two-month time scales.

\begin{table}[t!]
    \centering
    \begin{tabular}{lcc}
        \toprule
        Forecast horizon & Ratio & Preferred lag set \\
        \midrule
        30 days & 0.857 & [0,14,45] \\
        60 days & 0.806 & [0,14,45] \\
        90 days & 0.835 & [0,14,30,45] \\
        \bottomrule
    \end{tabular}
    \caption{
        Delayed-SINDy forecast skill relative to persistence for the three forecast horizons considered. The table reports the forecast ratio and delayed-state coordinates used by the sparse delayed-map model.
    }
    \label{tab:sindy_results}
\end{table}

Unlike ridge regression, which restricts the forecast map to a linear combination of delayed coordinates, delayed SINDy allows nonlinear polynomial interactions among the delayed coordinates while retaining only those terms required by the data. Delayed-SINDy forecasts outperform persistence at all three forecast horizons. The resulting models remain compact, requiring only a few delayed coordinates with memory scales of several weeks. These results support the broader picture emerging from the ridge and delayed-MLP analyses: much of the predictable variability resides in the reconstructed delayed state itself.

The 90-day forecast is particularly noteworthy because the sparse regression eliminates all nonlinear terms despite having access to a library containing polynomial interactions up to degree three. Using the delayed state associated with the lag set $\mathcal T=\{0,14,30,45\}$ days, the identified forecast map takes the form
\begin{equation}
    \widehat L(t_n+90)
    =
    0.1653\,L(t_n)
    +
    0.0477\,L(t_n-14)
    +
    0.0163\,L(t_n-30)
    +
    0.0653\,L(t_n-45).
\end{equation}
The resulting forecast depends only on the current Loop Current extension together with delayed values spanning the memory window identified in the previous analyses. The emergence of such a compact linear recurrence, despite allowing for nonlinear terms, indicates that much of the predictable variability is captured by the delayed-state representation itself rather than by increasing model complexity.

\begin{figure}[t!]
    \centering
    \includegraphics[width=.8\textwidth]{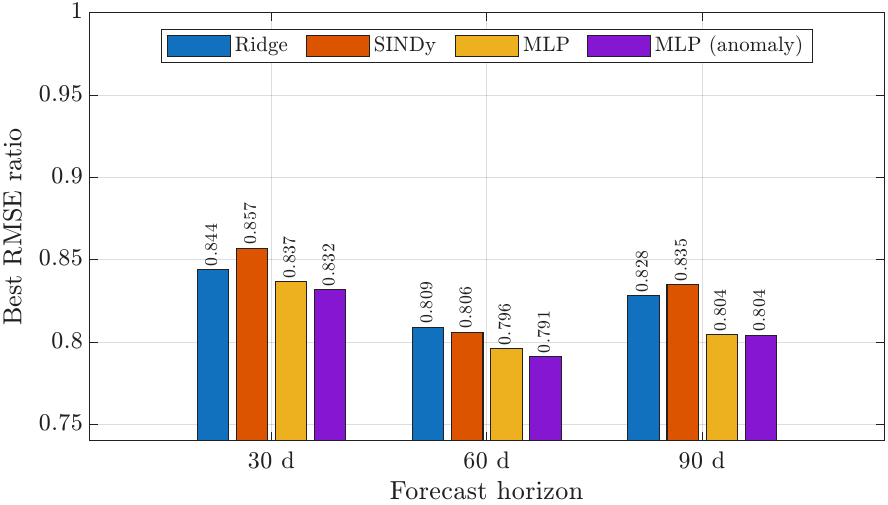}
    \caption{
        Comparison of ridge regression, delayed SINDy, delayed MLP, and anomaly-delayed MLP forecasts at 30-, 60-, and 90-day forecast horizons. Values indicate the best RMSE ratio relative to persistence.
    }
    \label{fig:lc_predictive_skill}
\end{figure}

The ridge, delayed-MLP, and delayed-SINDy forecasts all substantially outperform persistence. Figure~\ref{fig:lc_predictive_skill} compares the best forecast ratios obtained with each approach. The improvement relative to persistence is largest at intermediate lead times, where persistence has degraded while the delayed state still retains predictive information.

The overall differences among the methods remain modest. The similarity of the ridge, delayed-SINDy, and delayed-MLP forecast ratios suggests that much of the predictable variability is already contained in the delayed-state representation itself. This supports the view that the delayed state, rather than the specific forecasting model, provides the dominant source of predictive information. Nonlinear neural-network forecasts provide additional improvement, but the gains relative to ridge regression and delayed SINDy are comparatively small. The anomaly-delayed MLP results are nearly identical to those obtained using the original extension time series, indicating that the forecast skill is not primarily associated with the seasonal cycle or other slowly varying background variability.

\subsection{Recursive validation}

Accurate direct forecasts on a held-out test interval do not necessarily imply that the learned model defines a meaningful dynamical system. In the direct (open-loop) evaluations considered above, each forecast is initialized using delayed coordinates constructed from observations. The model therefore does not experience the effect of its own previous forecast errors.

Recursive (closed-loop) integration provides a more stringent test because forecast errors are allowed to accumulate through time. To assess dynamical consistency, the delayed-SINDy maps were initialized at the beginning of the test interval and then evolved autonomously using only their own previous predictions, with no observations supplied beyond the initialization time.

For a forecast horizon $\Delta$, the recursive prediction at a verification time $t_k$ is compared with the observed extension $L(t_k)$. The persistence forecast is evaluated at the same verification times according to
\begin{equation}
    \widehat L_{\rm pers}(t_k)
    =
    L(t_k-\Delta).
\end{equation}
Thus, the persistence curve corresponds to the observed record shifted by one forecast interval, whereas the recursive SINDy trajectory is generated autonomously from the initial delayed state. Because the Loop Current extension evolves slowly, persistence naturally resembles the observed time series, particularly at short forecast horizons. Forecast skill is therefore determined by the reduction in error relative to this shifted-observation baseline. Both forecasts are scored against the same observed values $L(t_k)$.

\begin{figure}[t!]
    \centering
    \includegraphics[width=.95\textwidth]{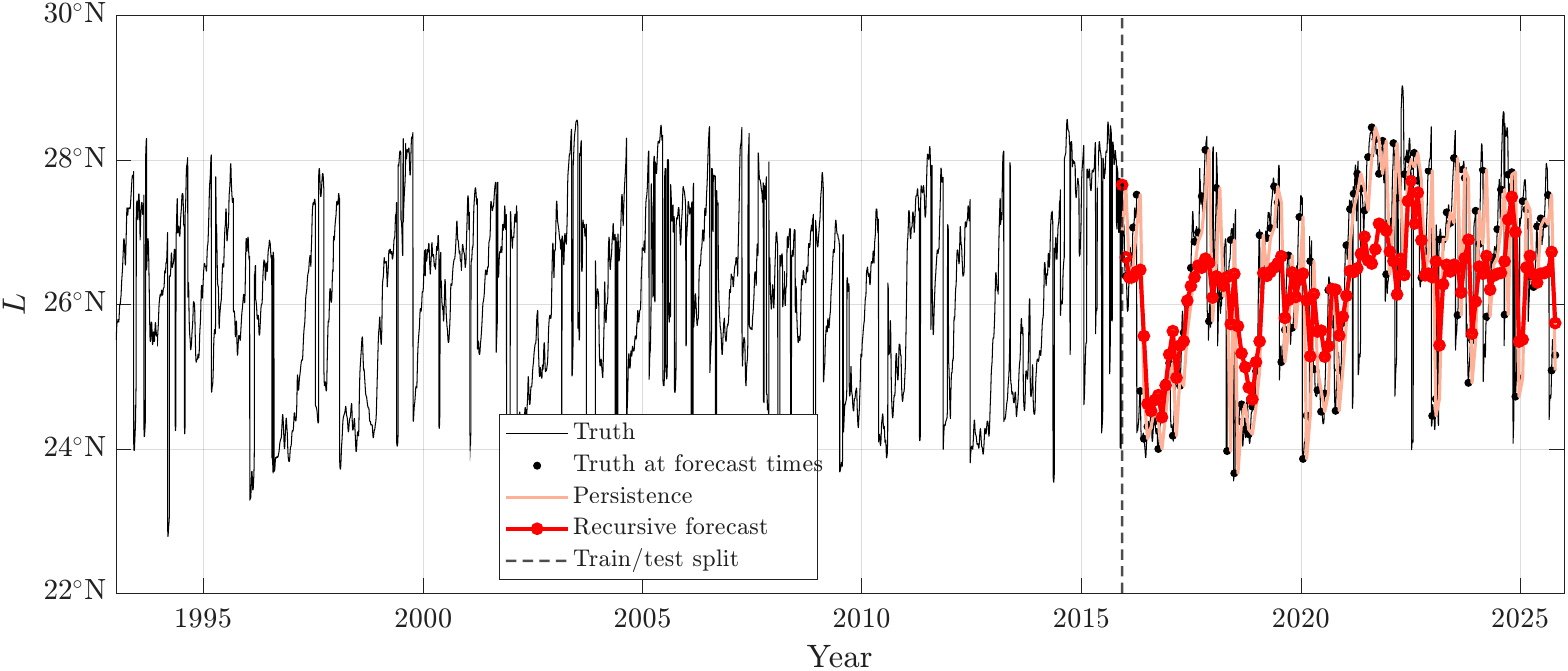}
    \caption{
       Autonomous recursive evolution of the 30-day delayed-SINDy map initialized at the beginning of the test interval. The persistence forecast is evaluated at the same verification times and corresponds to the observed extension shifted by the forecast horizon, $\widehat L_{\rm pers}(t_k)=L(t_k-\Delta)$.
    }
    \label{fig:lc_recursive_sindy_30d}
\end{figure}

Table~\ref{tab:recursive_validation} compares direct and recursive forecast skill. The recursive ratios remain close to the corresponding direct values at all forecast horizons. The largest change occurs for the 60-day forecasts, where the ratio increases from 0.806 to 0.833, while the 30-day and 90-day forecasts increase from 0.857 to 0.866 and from 0.835 to 0.851, respectively. These small differences indicate that forecast errors do not grow rapidly when the delayed maps are propagated recursively. Figure~\ref{fig:lc_recursive_sindy_30d} illustrates the recursive evolution of the 30-day delayed-SINDy map, which provides the highest temporal resolution example of autonomous propagation over the test interval.

\begin{table}[t!]
    \centering
    \begin{tabular}{lcc}
        \toprule
        Forecast horizon & Direct & Recursive \\
        \midrule
        30 days & 0.857 & 0.866 \\
        60 days & 0.806 & 0.833 \\
        90 days & 0.835 & 0.851 \\
        \bottomrule
    \end{tabular}
    \caption{
        Comparison between direct and recursive delayed-SINDy forecasts. Recursive forecasts are generated autonomously after initialization using only previous model predictions.
    }
    \label{tab:recursive_validation}
\end{table}

The recursive trajectories remain confined to the observed range of Loop Current extension and continue to reproduce the dominant low-frequency variability throughout the test interval. No evidence of instability, secular drift, or collapse is observed. The small differences between direct and recursive forecast skill indicate that the delayed maps remain dynamically coherent under autonomous iteration, supporting their interpretation as effective delayed evolution laws on the reconstructed state.

\section{Physical coordinates of the effective state}

The existence of a compact delayed representation raises the question of what physical variables parameterize the underlying state. In a Takens-type reconstruction, delayed observations can provide coordinates for the dynamics without necessarily identifying the physical processes represented by those coordinates. We therefore tested whether additional diagnostics associated with known mechanisms of Loop Current variability provide independent state information beyond the history of \(L(t)\).

The diagnostics considered here are motivated by earlier dynamical descriptions of the Loop Current. The models of Reid \cite{Reid-72} and Hurlburt and Thompson \cite{Hurlburt-Thompson-80} relate Loop Current extension to transport and potential-vorticity dynamics associated with the flow entering and exiting the Gulf of Mexico. More recently, Manta et al.~\cite{Manta-etal-23} investigated the predictability of Loop Current evolution from variability near the Yucatan Channel and identified velocity patterns correlated with future Loop Current states.

Motivated by these mechanisms, we considered transport estimates, transport proxies, mean speed, relative vorticity, and shear derived from satellite-altimetry velocities averaged over the Yucatan Channel and Florida Straits regions. We also considered geometric diagnostics measuring the distance between the Loop Current front and Cuba near the main gateway regions. Finally, cyclonic and anticyclonic vorticity indicators in the northern Caribbean Sea south of the Yucatan Channel were included as a simple measure of incoming mesoscale activity. These Eulerian vorticity diagnostics do not identify coherent vortex boundaries; a fully objective vortex analysis would require Lagrangian approaches such as geodesic eddy detection \cite{Haller-Beron-13,Haller-Beron-14}. 

The diagnostic comparisons used the same delayed-coordinate structure identified from the Loop Current extension analysis. This choice was made to isolate the effect of adding physical variables from changes associated with selecting a different memory structure. The objective was not to optimize a separate delayed representation for each diagnostic, but to test whether these quantities provide additional coordinates for the effective state reconstructed from \(L(t)\).

Forecast skill for the different predictor sets is summarized in Table~\ref{tab:transport_comparison}. The physical diagnostics contain predictive information when used independently, particularly at longer forecast horizons. However, adding these quantities to the delayed representation of \(L(t)\) does not systematically improve forecast skill.

\begin{table}[t!]
    \centering
    \begin{tabular}{lccc}
        \toprule
        Predictor set & 30 d & 60 d & 90 d \\
        \midrule
        Delayed $L(t)$ only & 0.878 & 0.801 & 0.847 \\
        Yucatan diagnostics only & 0.968 & 0.999 & 0.985 \\
        Florida Straits diagnostics only & 1.132 & 0.994 & 1.102 \\
        Gateway diagnostics only & 1.075 & 0.926 & 0.907 \\
        Caribbean vorticity diagnostics only & 1.058 & 0.916 & 0.945 \\
        \midrule
        Delayed $L(t)$ + Yucatan diagnostics & 0.903 & 0.905 & 1.051 \\
        Delayed $L(t)$ + Florida diagnostics & 0.978 & 0.972 & 1.036 \\
        Delayed $L(t)$ + gateway diagnostics & 0.931 & 0.942 & 0.890 \\
        Delayed $L(t)$ + Caribbean diagnostics & 0.955 & 0.912 & 0.944 \\
        \midrule
        Yucatan + Florida diagnostics & 1.069 & 0.964 & 1.075 \\
        Full predictor set & 0.989 & 1.014 & 0.990 \\
        Full predictor set + Caribbean diagnostics & 0.989 & 0.925 & 1.006 \\
        \bottomrule
    \end{tabular}
    \caption{
        Forecast skill relative to persistence for delayed Loop Current extension and additional physical diagnostics. Values below one indicate improvement over persistence.
    }
    \label{tab:transport_comparison}
\end{table}

These results do not imply that Yucatan Channel transport, Florida Straits variability, Caribbean mesoscale activity, or basin geometry are dynamically unimportant. Instead, they suggest that their influence is largely integrated into the recent evolution of the Loop Current front. Among the candidate state descriptions tested here, the delayed history of \(L(t)\) provides the most compact representation of the predictable variability.

This conclusion is limited to the physical diagnostics considered in this study. The effective low-dimensional state suggested by delayed reconstruction may involve other coordinates not examined here, including variables associated with subsurface circulation, potential-vorticity structure, or objective Lagrangian descriptions of the evolving Loop Current geometry. The present results show that the tested diagnostics do not provide substantial independent information beyond delayed observations of \(L(t)\), but they do not rule out alternative physical parameterizations of the reduced state.

A second consideration is observational. The transport and vorticity diagnostics are derived from satellite-altimetry products. While altimetry accurately captures large-scale Loop Current geometry, direct transport measurements or high-resolution numerical simulations would provide a more complete test of the role of inflow and outflow variability.

The effective low-dimensional state identified by delayed reconstruction is therefore not simply parameterized by the additional physical diagnostics considered here. Rather, Loop Current extension appears to act as a coarse-grained observable whose history captures much of the dynamically relevant information.

\section{Relation to previous delayed-coordinate analyses}

Earlier studies have suggested that Loop Current variability may admit a reduced dynamical description. In particular, Lugo-Fern\'andez \cite{Lugo-07} used delayed-coordinate reconstruction and nonlinear time-series analysis to investigate whether Loop Current variability exhibits signatures of low-dimensional deterministic dynamics.

The present work revisits this question from a forecasting and model-discovery perspective. Rather than using delayed-coordinate reconstruction to estimate attractor properties, such as embedding dimension or measures of deterministic structure, we ask whether delayed coordinates define a state representation capable of supporting predictive evolution maps. Ridge regression is used to measure how much forecast-relevant information is contained in delayed observations of Loop Current extension, while delayed MLP and delayed SINDy provide nonlinear and sparse representations of the resulting delayed dynamics.

The results are consistent with the existence of a compact effective state controlling Loop Current extension variability. Forecast skill saturates after only a few delayed coordinates, with preferred memory scales near 14--45 days. In addition, delayed SINDy maps retain skill under recursive propagation, indicating that the reconstructed coordinates support an effective delayed evolution law rather than only isolated forecasts.

These results complement earlier delayed-coordinate studies by connecting reconstruction of Loop Current variability with prediction and data-driven model discovery. The focus is not only whether low-dimensional structure is present, but whether such structure can be used to construct compact evolution maps.

\section{Concluding remarks}

The Loop Current is often described through its cycles of extension, retraction, and eddy shedding, yet it remains unclear how much of this variability can be captured by a reduced dynamical description. The results presented here indicate that a substantial fraction of the predictable variability is contained in a compact delayed representation of Loop Current extension itself.

Several conclusions emerge:
\begin{enumerate}[label=(\roman*),nosep]
    \item Forecast skill saturates after only a few delayed coordinates, with dominant memory scales of approximately two weeks to one-to-two months.
    \item Ridge regression, delayed MLP, and delayed SINDy achieve broadly comparable performance, indicating that much of the predictive information resides in the reconstructed state rather than in the particular forecasting model.
    \item Delayed SINDy identifies compact delayed evolution maps that retain forecast skill under recursive propagation and reproduce the dominant low-frequency variability over nearly a decade of testing.
    \item Physical diagnostics associated with the Yucatan Channel, Florida Straits, gateway geometry, and northern Caribbean vorticity contain predictive information but do not provide additional skill once the delayed Loop Current state is included.
\end{enumerate}

The similarity of the ridge, delayed-SINDy, and delayed-MLP forecasts suggests that increasing model complexity provides only modest gains beyond the information contained in the delayed coordinates. The delayed representation therefore emerges as the central object of interest. The diagnostic comparisons further suggest that the tested transport, geometric, and mesoscale indicators do not provide substantial independent state information beyond that already encoded in the recent evolution of the Loop Current front. This does not exclude the possibility that other physical variables provide more direct coordinates of the underlying state. In particular, high-resolution ocean simulations and longer observational records could allow additional diagnostics, including subsurface circulation and potential-vorticity structure, to be tested over a broader range of Loop Current regimes.

The identification of a compact delayed state may also prove useful for future forecasting systems. In practice, this suggests augmenting more sophisticated architectures with delayed observations spanning the dominant memory windows identified here, roughly two-week to one-to-two-month time scales, rather than treating the input history length as a purely tunable parameter. Whether such explicit delayed representations can improve performance or interpretability remains a topic for future work.

The present analysis defines Loop Current extension using a calibrated Eulerian contour of absolute dynamic topography. This provides a simple and reproducible observable for studying delayed predictability, but it is not intended to define an objective material boundary of the Loop Current. Future work could combine delayed-state modeling with Lagrangian descriptions of the evolving circulation geometry. For example, shearless transport barriers and geodesic eddy diagnostics provide observer-independent methods for identifying coherent structures in unsteady flows. Such approaches could be used to investigate whether similar low-dimensional delayed dynamics arise when the Loop Current state is defined from material transport structures rather than Eulerian contours.

The available record remains relatively short when viewed against the full range of Loop Current variability. Longer observations would provide a more stringent test of the stability of the preferred lag sets, delayed evolution maps, and recursive forecasts across different dynamical regimes. The results nevertheless support the view that Loop Current extension behaves as an effective delayed dynamical system whose predictable variability can be reconstructed from a small number of delayed observations and represented through compact delayed evolution maps.

\section*{Acknowledgments}

The authors thank Gage Bonner for many helpful discussions on SINDy.

\section*{Funding}

Early development of this work was supported by the Gulf Research Program of the National Academies of Sciences, Engineering, and Medicine under award UGOS \#2000011056.

\section*{Author Declarations}

\subsection*{Conflict of Interest}

The authors have no conflict of interest to disclose.

\section*{Author contributions}

F.J.B.-V.\ conceived the study, developed the methodology, performed the numerical analysis, generated the figures, and wrote the manuscript. M.J.O.\ co-conceived the study and contributed to manuscript review and editing. P.M.\ contributed to the numerical analysis and to manuscript review and editing. ChatGPT assisted with code development, debugging, and manuscript editing.

\bibliographystyle{alpha}
\bibliography{fot}

\section*{Data and Software Availability}

The gridded multimission altimeter products were produced by SSALTO/DUACS and distributed by AVISO (\href{https://www.aviso.altimetry.fr/}{https://www.aviso.altimetry.fr/}) with support from CNES.

The MATLAB scripts used to generate the figures and results presented in this paper are available from the corresponding author upon reasonable request.

\end{document}